# The Economic End of Life of Electrochemical Energy Storage

Guannan He[a,b,c], Rebecca Ciez[d], Panayiotis Moutis[e], Soummya Kar[e], and Jay Whitacre[a,b,f]*

**Abstract**
The useful life of electrochemical energy storage (EES) is a critical factor to system planning, operation, and economic assessment. Today, systems commonly assume a physical end-of-life criterion: EES systems are retired when their remaining capacity reaches a threshold below which the EES is of little use because of insufficient capacity and efficiency. We have found, however, that there are some instances where, while the EES is still functional, it is no longer economically profitable; we call this criterion the economic end of life of the system. This criterion depends on the use case and degradation characteristics of the EES. Using an intertemporal operational framework to consider functionality and profitability degradation, our case study shows that the economic end of life could occur significantly faster than the physical end of life. We argue that both criteria should be applied in EES system planning and assessment. We also analyze how R&D efforts should consider cycling capability and calendar degradation rate when considering the economic end of life of EES.

## 1    Introduction

Nearly all future energy technology assessments find that distributed and/or centralized electrochemical energy storage (EES) with favorable economics in particular, is essential to enabling a clean, sustainable, and low-carbon energy future[1-5]. The degradation behavior of EES is a critical component to assessing its economic viability: as EES ages, available capacity fades and internal impedance rises due to various side-reactions.

Previous studies have developed methods that model[6-10] and monitor[11-13] the degradation process and predict the EES life[14-19] to inform investment, planning, and operation decisions[20-26]. Typically, the functionalities (including energy capacity, power capacity, energy efficiency, etc.) of EES degrade as energy is processed and as time goes by[7,9,21,27-29]. While the ageing processes can take place in any sub-components of the EES technology, the observed losses in function/capability are commonly due to a combination of interacting degradation mechanism[30]. The capacity loss in lithium-ion batteries typically occurs when the electrode materials becomes coated by a solid electrolyte interphase (SEI) that both consumes lithium and limits ion transport[31,32]. SEI formation also results in contact loss in the composite anode, which leads to impedance increase and in turn, power fade and efficiency fade[15,28,30,33].

We can divide the degradation of most EES cells into two categories: 1) cycling degradation that mainly depends on the amount of energy throughput the cell has processed; 2) calendar degradation that mainly depends on the length of time the EES has experienced. The cycling degradation is usually modelled as a function of cycle number, depth of discharge (DOD), and temperature, while the calendar degradation is modelled as a function of time, state of charge (SOC), and temperature[27,28].

[a] Department of Engineering and Public Policy, Carnegie Mellon University, Pittsburgh, PA 15213, USA
[b] Wilton E. Scott Institute for Energy Innovation, Carnegie Mellon University, Pittsburgh, PA 15213, USA
[c] MIT Energy Initiative, Massachusetts Institute of Technology, Cambridge, MA 02139, USA
[d] Andlinger Center for Energy and the Environment, Princeton University, Princeton, NJ 08544, USA
[e] Department of Electrical Engineering, Carnegie Mellon University, Pittsburgh, PA 15213, USA
[f] Department of Materials Science and Engineering, Carnegie Mellon University, Pittsburgh, PA 15213, USA
(Email: whitacre@andrew.cmu.edu)

To assess the cumulative degradation of EES, the state of health (SOH), as an indicator, is most commonly implemented in both academia and industry. It is defined as the ratio of the remaining capacity to the initial rated capacity. For example, if the initial capacity is 100 kWh and the current capacity is 95 kWh, then the current SOH is 95%. Thus far, the end of life (EOL) of EES has been determined by some physical criteria, e.g., when the SOH decreases to 80% or 70% [7,27,34]. This physical criterion of EOL is not rigorous—the EES may still be usable after the SOH reaches 70%, and different amounts of energy may be available depending on the discharge currents used to assess battery SOH. However, it is possible, depending on degradation mode, that a more sudden "death" will occur after the physical EOL, which means that the capacity will decrease and the impedance will increase at a much more drastic rate[27]. Safety may also be compromised when using the EES after its physical EOL.

The physical EOL criterion does not account for concurrent reductions in profitability, and eventually EES profits may be insufficient to compensate for fixed operating and maintenance (O&M) costs such as land rent, property tax, insurance expense, labor cost, etc[35]. If this occurs before the physical life of EES ends, the initial life prediction would have been over-optimistic, and the EES operation would have to be terminated earlier than expected or be augmented. Moreover, the indicator SOH that only reflects the energy capacity is too simplistic for EES profitability estimation. The power capacity fade and efficiency decreases are also determining factors and could be even more critical than the energy capacity to profitability reduction.

In this article, we explore the novel concept of an economic EOL for EES that considers whether the EES operation should be either terminated or substantially altered for financial reasons prior to physical system failure. We do this using an intertemporal operational decision framework which maximizes the life-cycle benefit of EES considering functionality and profitability degradation[36]. In the framework, an opportunity cost (named as the marginal benefit of usage) that reflects the future use value of EES is calculated and incorporated into EES operational decisions. Various degradation in functionality including energy capacity, power capacity, and energy efficiency are considered in the operational decisions. Given the optimal operational decisions with the maximal life-cycle benefit, we calculate the cash flow of the EES project and determine the economic EOL. Because the economic EOL criterion does not depend on initial capital costs, instead focusing entirely on the remaining profitability of the system, this criterion is applicable to both new EESs and secondary EESs retired from electric vehicle and reused for grid applications[37-39].

The case study results indicate that the economic life of EES decreases from utility to commercial and residential applications, because the economic life decreases as the fixed O&M cost increases, while fixed O&M cost depends on EES size and application. We further analyze how the cycling capability and calendar degradation rate affect overall profitability of EES in energy arbitrage application using the economic end-of-life criterion.

## 2 Methods

### 2.1 Intertemporal operational framework

To simulate the operational decisions of EES and evaluate the cash flow over its life cycle, we implement an intertemporal operational framework that maximizes the net life-cycle benefit of EES (initial capital cost is not accounted as it is already incurred at the operational stage). The model we use in this study is improved from our previous work[36] by dynamically updating power and energy capacity and energy efficiency that significantly affect the EES profitability. The marginal benefit of usage (MBU) is introduced to bridge the short-term, mid-term and long-term. Given forecasting information, the model is proved to produce the optimal short-term outputs with the maximum net life-cycle benefit subject to degradation constraints. The mathematic formulations of the model are:

$$\text{LB}_{\max} = \max_{\mu} \ \text{LB} = \max_{\mu} \ \sum_{t \leq T} \delta_t \text{SB}_t(\mu) \tag{1}$$

$$\text{s.t.} \quad \sum_{t \leq T} d_t(\mu) \leq D \tag{2}$$

$$d_t(\mu) \geq C \tag{3}$$

$$\text{SB}_t(\mu) = \max_{\mathbf{P}_t \in \mathbf{F}(\text{SOH}_t)} r_t(\mathbf{P}_t) \tag{4}$$

$$\text{s.t.} \quad \mathbf{P}_t = \mathbf{0} \ \text{ or } \ \frac{\partial \text{SB}_t(\mu)}{\partial d_t(\mathbf{P}_t)} = \frac{\mu}{\delta_t} \tag{5}$$

In the long term, the EES operator maximize the net life-cycle benefit by determining the optimal life-cycle MBU $\mu$ subject to EES degradation constraints, as Equations (1)-(3). $\text{LB}$ represents the life-cycle benefit of EES; $\text{SB}_t$ is the maximum short-term benefit at time $t$ as a function of the EES degradation at time $t$, denoted by $d_t$, respectively; $D$ is the degradation (energy throughput) limit over the EES lifetime or the remaining energy throughput for an old EES; $T$ is the length of the EES lifetime determined by the EES degradation rates $d_t$ and the degradation limit $D$; $C$ is the calendar degradation rate. The net life-cycle benefit is calculated by aggregating all simulated net mid-/short-term benefits, as Equation (1). Equation (2) models that the total usage/energy throughput over the EES life has a limit, determined by the physical end-of-life criterion. Equation (3) models the calendar degradation of the EES system.

In the mid-term, typically a horizon between a month and a year, the EES operator updates the discounted MBU $\frac{\mu}{\delta_t}$ by multiplying the life-cycle MBU by a discounting factor. The discounted MBU combines long-term information and the time preference of EES operator on profits earned in different periods—it is always preferable to earn money sooner than later, and thus the EES should be utilized more in earlier years over its life—and feeds into the short-term to guarantee that the life-cycle benefit is maximized when short-term decisions are made.

In the short term, typically a day, the EES operator determines the optimal short-term outputs (hourly or intra-hourly) to maximize the daily benefits, based on the current SOH, the forecasted market prices, and the discounted MBU updated in the mid-term, as Equations (4)-(5). $r_t(\mathbf{P}_t)$ is the short-term benefit at time $t$ as a function of the charge/discharge schedules at time $t$ (denoted as $\mathbf{P}_t$); and $\mathbf{F}(\text{SOH}_t)$ is the feasible operating set of the EES, typically convex, and is a function of the SOH at time $t$; The EES degradation at time $t$, $d_t$, can also be expressed as a function of the charge/discharge schedules $\mathbf{P}_t$. The role of the discounted MBU is similar with a marginal cost per unit of degradation but should be interpreted as the required marginal benefit per unit of degradation. Equation (5) is a necessary optimality condition (first-order Karush–Kuhn–Tucker condition) of the long-term maximization problem (Equations (1)-(3)): if the EES is not operating, the short-term solution becomes trivial with $\mathbf{P}_t = \mathbf{0}$ and $d_t = C$; if the EES is operating, the derivative of the short-term objective in terms of the usage/degradation should be equal to the discounted MBU.

Please see the Electronic Supplementary Information for a more detailed explanation of how to implement the intertemporal operational framework and how the cash flow and the economic end of life are calculated.

Given the maximum net life-cycle benefit $LB_{max}$ and the usage limit over the EES lifetime, the average benefit of usage is calculated as:

$$ABU = \frac{LB_{max}}{D} \tag{6}$$

### 2.2 Short-term energy arbitrage model

The short-term decision optimization model is presented as Equations (7)-(12):

$$SB_t = \max_{P_h^{dis}, P_h^{cha}} \sum_{h \in [t, t+\Delta t]} \lambda_h^e (P_h^{dis} - P_h^{cha}) - c_{fix} \tag{7}$$

$$\text{s.t.} \quad \frac{\partial SB_t}{\partial d_t} = \frac{\mu}{\delta_t} \quad \text{if} \quad d_t > C \tag{8}$$

$$d_t = \sum_{DOD} 2E_t^{max} n_{t,DOD} DOD^k + C \tag{9}$$

$$E_h = (1-\rho)E_{h-1} + P_h^{cha}\eta_t(SOH_t) - P_h^{dis}/\eta_t(SOH_t) \tag{10}$$

$$0 \leq P_h^{dis}, P_h^{cha} \leq P_t^{max}(SOH_t) \tag{11}$$

$$0 \leq E_h \leq E_t^{max}(SOH_t) \tag{12}$$

The short-term (daily) benefit, as in equation (7), is the sum of revenues at each hour $h$ within the time interval $[t, t+\Delta t]$, minus the fixed O&M cost $c_{fix}$. $n_{t,DOD}$ denotes the number of cycles at certain DOD during $[t, t+\Delta t]$, and is determined by the discharging and charging schedules, $P_h^{dis}$ and $P_h^{cha}$. In equation (9), $E_t^{max}$ is the EES energy capacity during time $t$, and $2E_t^{max}$ represents the energy throughput of a full cycle, including both charging and discharging. For energy arbitrage, EES typically takes one or two cycles per day, and thus we can estimate the degradation by setting $n_{t,DOD} = 1$ and $DOD = \sum_{h \in [t,t+\Delta t]} \frac{P_h^{dis} + P_h^{cha}}{2E_t^{max}}$. Equation (10) describes the charging/discharging process of the EES as a function of its state of charge (SOC), where $E_h$ is the SOC at hour $h$; $\rho$ is the self-discharge rate; $\eta_t$ is the charge/discharge efficiency during time $t$. Equations (11) and (12) indicate the physical constraints of the power output and the SOC of the EES, where $P_t^{max}$ is the EES power capacity during time $t$. The energy and power capacity and charge/discharge efficiency are all functions of the SOH at time $t$. Since the SOH decrease over one short-term period is tiny, we can assume the SOH is constant over period $t$ and is equal to the SOH at the beginning of the period.

### 2.3 Simulation procedures

The main task of the intertemporal operational decision framework is to find the optimal marginal opportunity cost, named as MBU, that leads to the maximum life-cycle benefit subject to degradation constraint.

First, we need to simulate short-term operations of EES to calculate the short-term benefits given different

MBUs, as the first step in Fig. 1(a). Let's assume the time horizon of short term is a day. To estimate the daily benefit given an MBU, we solve the short-term decision model, which include Equations (7)-(12) for energy arbitrage application. The objective is to maximize the daily benefit, given price forecasts and the EES health status, and the decision variables are the EES daily outputs, as shown in Fig. 1(b). One possible way to solve the model is to first exclude constraint (8) and solve the remaining model for each discrete levels of daily degradation (setting equation (9) to different values in a reasonable set). Then we can find which degradation level satisfies constraint (8). The optimal outputs and benefit with the degradation level that satisfies constraint (8) is the final optimal solution to the short-term decision model. Given the optimal outputs and degradation, the state of health and the functionality of EES are updated for the next-day operation simulation. By repeatedly running the short-term decision model and changing the MBU value, we can obtain different series of short-term benefits over the EES life at different MBUs.

The second step, as shown in Fig. 1(a), is to calculate the discounted annual benefits and the life-cycle benefits for each MBUs by discounting and aggregating the short-term benefits within the EES life for each MBU, which is formulated as the summation in Equation (1) and constraint (2).

Step 3 in Fig. 1(a) is to find the optimal MBU that generates the maximum life-cycle benefit. With the optimal MBU, we can then obtain the corresponding optimal cash flow, a series of annual benefits.

## 2.4 Parameter setting

As a baseline, we assume approximate degradation characteristics that, while not perfectly accurate for all batteries, are reasonable for the purposes of our analysis. Specifically, we assume that after 3000 charge-discharge cycles at 100% DOD or 30000 cycles at 10% DOD, the remaining energy capacity of the EES decreases to 70% of the original capacity and the impedance increases to 200% of the original; the number of cycles to physical failure is a power function of the DOD (with a power of -1)[40]; the calendar degradation of the EES is 1% capacity loss/year[21,27,29]; the original charge/discharge efficiency is 90%; and the original energy to power ratio is 4 hours. The effect of any stress event, cycling or calendar, is assumed to be only dependent on the current state of EES and can be linearly accumulated[7]. We use the day-ahead energy prices of CAISO in 2016 to represent the price scenarios of each year during the EES lifetime, in which the average peak-valley price difference is approximately $32/MWh. Without loss of generality, the EES operator is assumed to be a price-taker, whose actions in the markets have no impact on the market equilibrium and prices. We use a discount rate of 7%, as a recommended value for private investment by the Office of Management and Budget of US[1]. The annual fixed O&M costs are 9, 16, and 27 $/kW-year for utility-scale, commercial/industrial, and distributed/residential lithium-ion EES, respectively[35]. This cost difference may result from the economics of scale—larger systems could have lower rent, insurance, and labor expenses per unit of installed capacity.

(a)

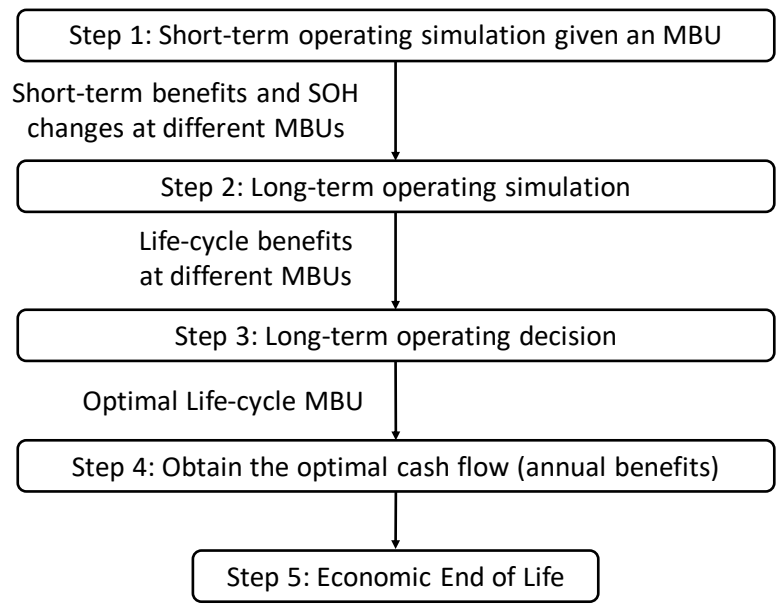

(b)

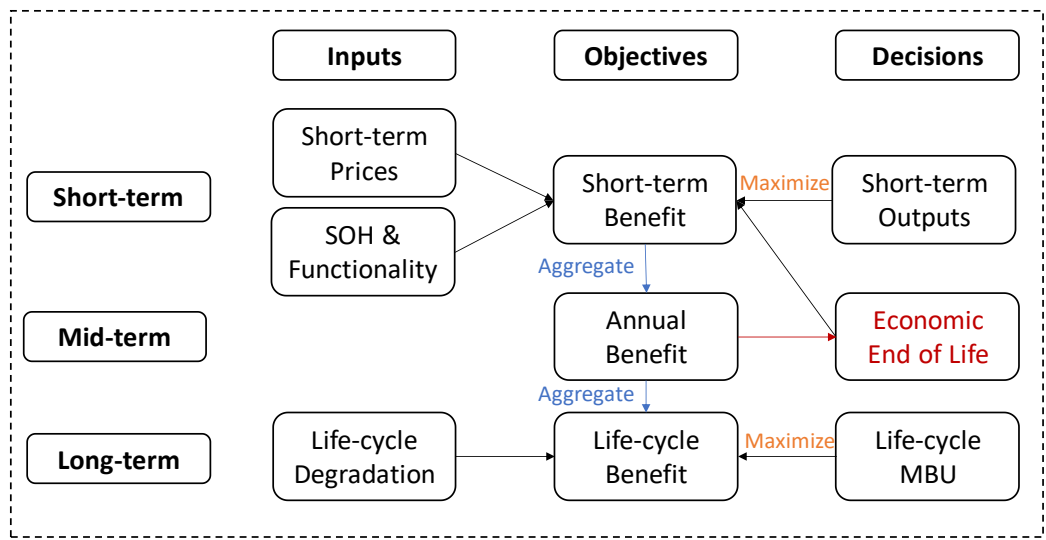

Figure 1. Schematic of the intertemporal operational decision framework to determine the economic end of life of EES. (a) The data-flow diagram shows the necessary procedures of simulations and decisions and data flow from each procedure to the next for the calculation of economic end of life. MBU: marginal benefit of usage. (b) Decision variables, objectives and key inputs for each decision and their relations.

## 2.5 Power capacity and energy efficiency estimation

We estimate the power capacity and energy efficiency of EES over time based on the internal impedance by assuming a series circuit model of the EES, which consists of a constant voltage source, the internal impedance, and a flexible external impedance. The constant voltage source and the internal impedance represent the EES. Denoting the original and current internal impedance by $Z_0$ and $Z_t$, respectively, and let $U$ and $I$ be the voltage and current on the external impedance, the current energy efficiency $\eta_t$ can be derived as

$$\eta = \frac{UI}{UI + I^2 Z_t} = \frac{1}{1 + Z_t \frac{I}{U}} \tag{13}$$

Let $\eta_0$ be the initial energy efficiency, we have:

$$\eta_0 = \frac{1}{1 + Z_0 \frac{I}{U}} \Leftrightarrow \frac{I}{U} = \frac{1 - \eta_0}{\eta_0 Z_0} \tag{14}$$

Both the initial and the current instantaneous energy efficiency are dependent on the external voltage and current. Assuming the same charge/discharge cycle with identical voltage and current curves, we can plug (14) into (13):

$$\eta_t = \frac{1}{1 + \frac{Z_t}{Z_0} \frac{1 - \eta_0}{\eta_0}} \tag{15}$$

For the power capacity, we assess the maximum external power that can be achieved given the constant voltage source and the internal impedance rather than assuming the same charge/discharge cycle. Denote the source voltage by $U_0$, we can easily derive the maximum power as:

$$P_t^{max} = \frac{U_0^2}{4 Z_t} \tag{16}$$

Let $P_0^{max}$ be the initial power capacity, we have:

$$P_0^{max} = \frac{U_0^2}{4 Z_0} \Leftrightarrow U_0^2 = 4 P_0^{max} Z_0 \tag{17}$$

By plugging (17) into (16), we have:

$$P_t^{max} = \frac{Z_0}{Z_t} P_0^{max} \tag{18}$$

# 3 Results

## 3.1 Economic EOL

We define the economic EOL for EES as the point in time beyond which the EES is unable to earn positive net economic benefit in its application. In those cases where the economic EOL is found to be earlier than the physical EOL, the EES should be retired and either recycled or put into a different economically favorable application.

As an example, we evaluate the economic EOL of a grid-tied lithium-ion EES system arbitraging in California. In the energy arbitrage application, the EES discharges/sells energy when the market price is high, while charges/buys energy when the market price is low. In some cases, when the abundancy of renewable energy determines the energy price level, energy arbitrage can also be interpreted as renewable energy integration. Although the economic EOL of applications other than energy arbitrage may have different properties, we focus on energy arbitrage in this paper as it is a promising solution to a low-carbon energy system. Because the fixed O&M cost is usually an annual mean value, the scale of the economic EOL is also a year.

The cash flows with different fixed O&M cost for utility-scale, commercial, and distributed EES are presented in Fig. 2, which correspond to the optimal operational decisions that maximize the life-cycle benefits of EES and are calculated using an intertemporal operational framework (see Method)[36]. We can see the net annual profits decrease with time for all three cases. This trend of decreasing annual profits is a direct result from both the functionality degradation and the temporal discounting of EES owner/operator on profits. In the case of utility-scale EES with an annual fixed O&M cost of $9/kW-year, the physical life of the EES ends in Year 8 with positive net annual profit, as in Fig. 2(a). In the case of commercial/industrial EES with an annual fixed O&M cost of $16/kW-year, however, the net annual profit becomes negative from Year 6, as in Fig. 2(b), because the annual profit is less than the annual fixed O&M cost. As such, Year 5 is the economic EOL by our definition and is earlier than the physical EOL. Therefore, it is very likely that the EES owner will find the economically useful life of an EES system ends earlier than the technical EOL claimed by the manufacturer. For distributed/residential EES, the economic EOL is Year 1 (Fig. 2(c)), which implies that the distributed/residential EES should only be applied where profit opportunities are much higher than arbitraging in the current wholesale energy markets, such as in providing frequency regulation, uninterruptible power supply, and so on. It is worth noting that the economic EOL is independent of the initial capital cost of EES—in fact, no matter how future cost of EES decreases, the economic EOL could still dominate the physical EOL for energy arbitrage.

The values of the fixed O&M cost used above may not be accurate, as they vary across different regions and projects. Fig. 3 presents the relationship between the ends of life and the annual fixed O&M cost of EES. The economic life of EES ranges from 11 years to 1 year as the annual fixed O&M cost increases from 0 to $30/kW-year. When the annual fixed O&M cost is larger than $12/kW-year, the economic EOL is earlier than the physical EOL, which implies that the economic EOL should be the true end of the EES project, if no replacement. In Fig. 3, the physical EOL also varies with different fixed O&M costs, since the corresponding optimal operational strategies are different. For example, if the fixed O&M cost increases, utilizing the EES more heavily in the early years of the EES life could be profitable, which accelerates EES degradation, shortens EES physical/economic life, and in turn saves some fixed O&M cost.

(a)

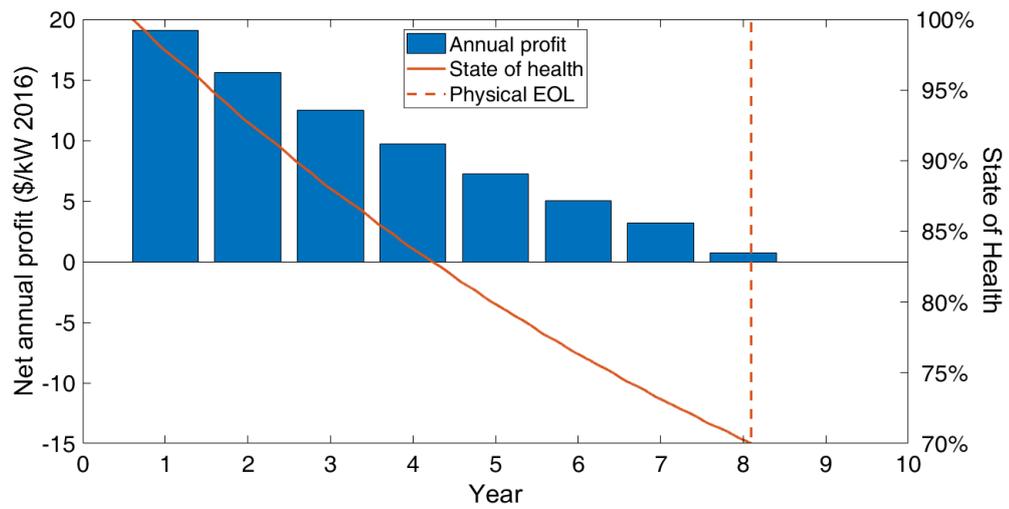

(b)

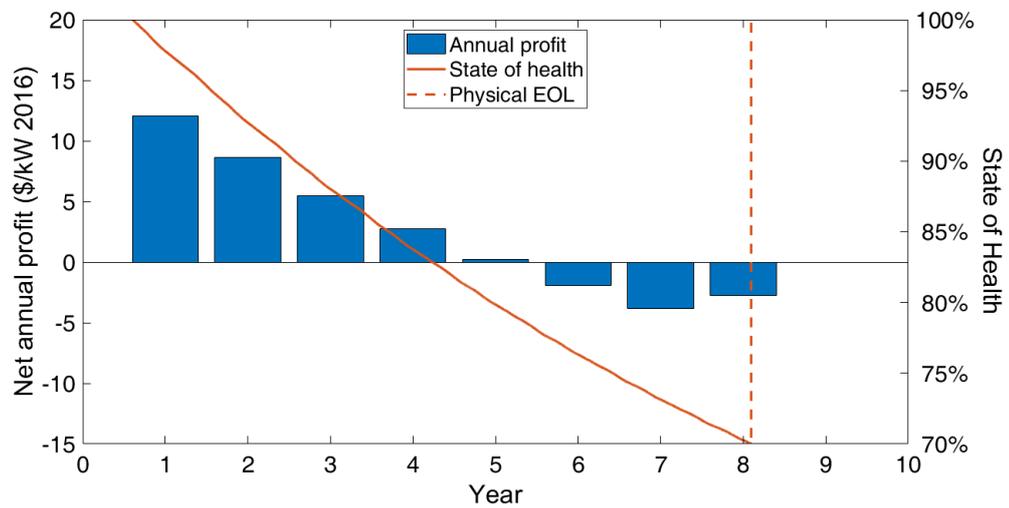

(c)

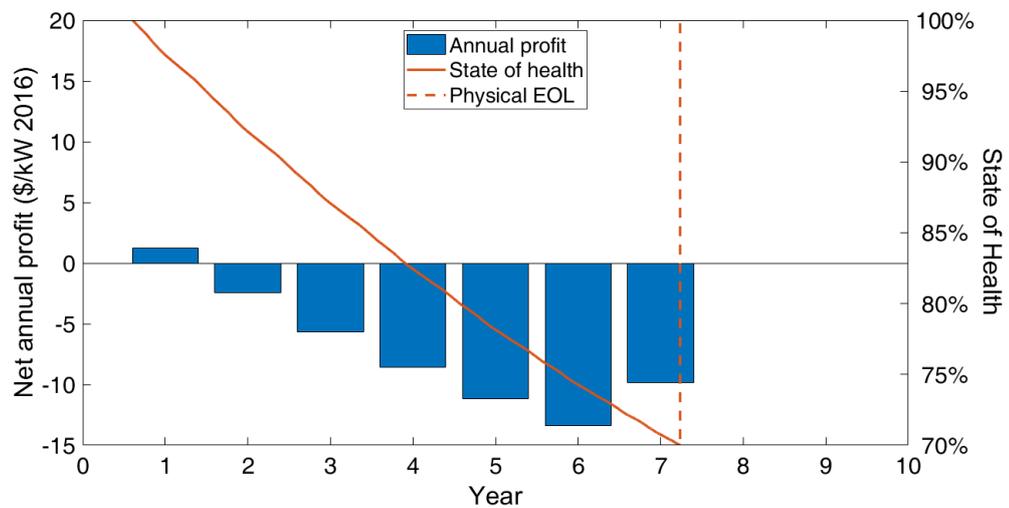

Figure 2. (a) Utility-scale storage with a fixed O&M cost of $9/kW-year. (b) Commercial/industrial storage with a fixed O&M cost of $16/kW-year. (c) Distributed/residential storage with a fixed O&M cost of $27/kW-year.

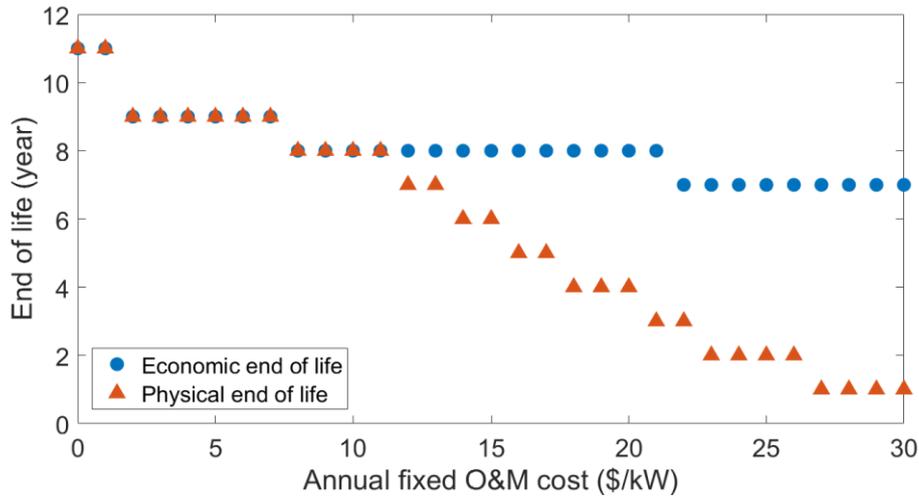

Figure 3. The changes of economic and physical ends of life with annual fixed O&M cost. Higher O&M costs incentivize use cases that degrade the EES more rapidly, resulting in a shorter physical lifetime.

### 3.2 Profitability and functionality fade

One major reason that the net annual profit is decreasing is the profitability and the functionality of EES degrade as more energy is processed and time goes by. Fig. 4 present how the profitability of EES measured in annual gross profit of a certain year and the functionalities including energy and power capacities and efficiency as SOH decreases. If the impedance rises to 200% after the SOH decreases to 70%, the power capacity decreases to 50% and the charge/discharge efficiency decreases from 90% to 82% (81% to 66% for round-trip efficiency) approximately. Energy efficiency is important for the profitability of EES in energy arbitrage application, as lower efficiency increases charging cost, while the power capacity is also critical to capturing the arbitrage opportunity in a day, e.g. the hours with the peak and valley prices. In this case, the annual gross revenue decreases from $30/kW to $24/kW approximately, a 20% loss in profitability, as SOH decreases from 100% to 70%.

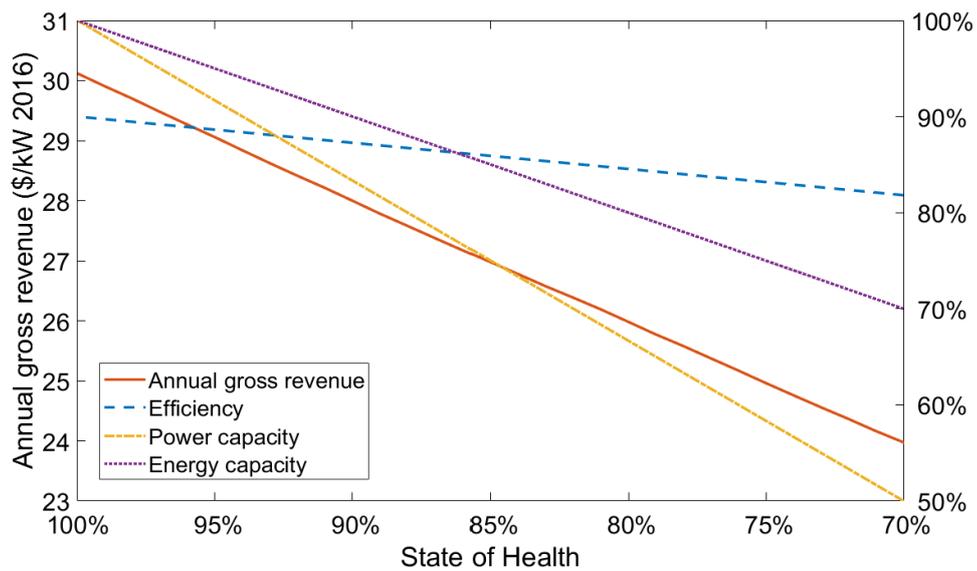

Figure 4. The changes of profitability and functionality of EES with SOH. The percentages on the right y-axis represent the ratios of the remaining capacity to the original capacity for power and energy capacity (yellow and purple lines). For efficiency (blue line), the percentages represent the actual values.

## 3.3 Degradation rate sensitivity

Figs. 4 and 5 depict how the life-cycle profitability of EES changes with the degradation rates using the economic EOL criterion. All parameters are the same as before except for the cycling capability (total number of 100%-DOD cycles before the physical EOL) and the calendar degradation rate, and the fixed O&M cost is $9/kW-year.

From Fig. 5, we can see that the unit-capacity profit of EES increases as the cycling capability increases and the calendar degradation rate decreases, as expected. If we look at the gaps between each contour, we can also see that as the calendar degradation rate increases, the marginal increase rate of unit-capacity profit with respect to improving cycling capability is decreasing—it requires more cycling capability improvement to increase one unit of unit-capacity profit at higher calendar degradation rate. This implies that improving cycling capability is a less efficient way to enhance the economic viability of EES when the calendar degradation rate is high, and it is unwise to emphasize the cycling capability improvement in R&D while disregarding reductions to the calendar degradation rate.

In comparison with Fig. 5, the profitability in Fig. 6 is in terms of the average benefit of usage, which is calculated by dividing the maximum total life-cycle benefit/profit by the total usage/energy throughput (charge plus discharge) over the EES life (see Equation (6) in Methods section). The average benefit of usage increases as calendar degradation rate decreases, while the average benefit of usage decreases as the cycling capability increases. The latter implies a diminishing marginal benefit of usage/energy throughput, which may disfavor some batteries with high cycling capability but also high cost.

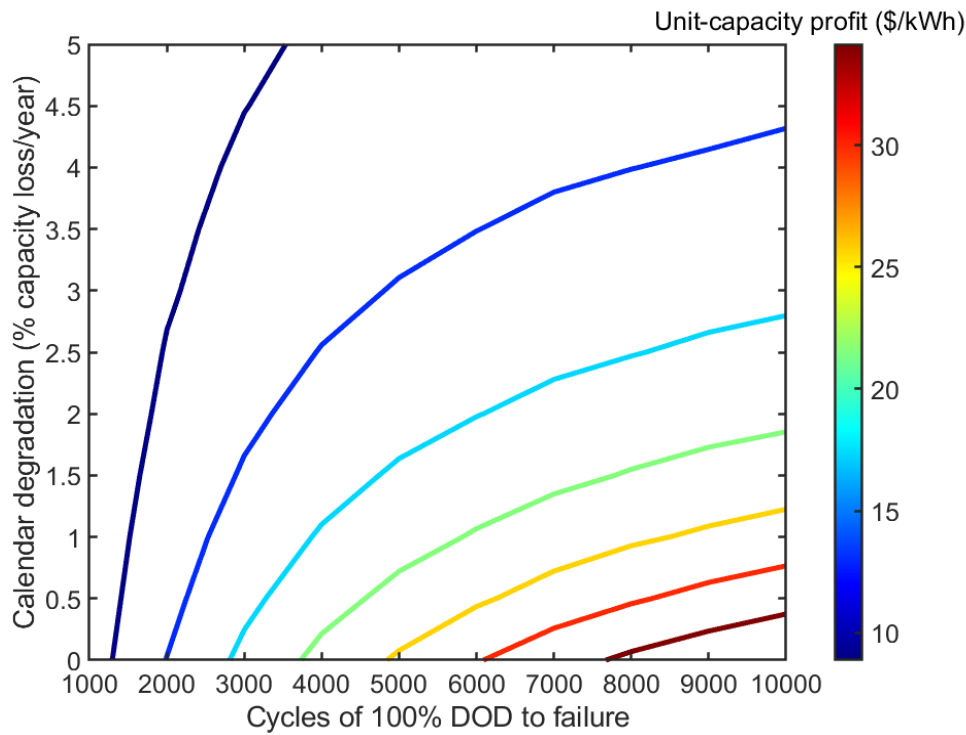

Figure 5. Sensitivity analysis on how the unit-capacity profit of EES arbitraging in California vary as the cycling and calendar degradation rates change

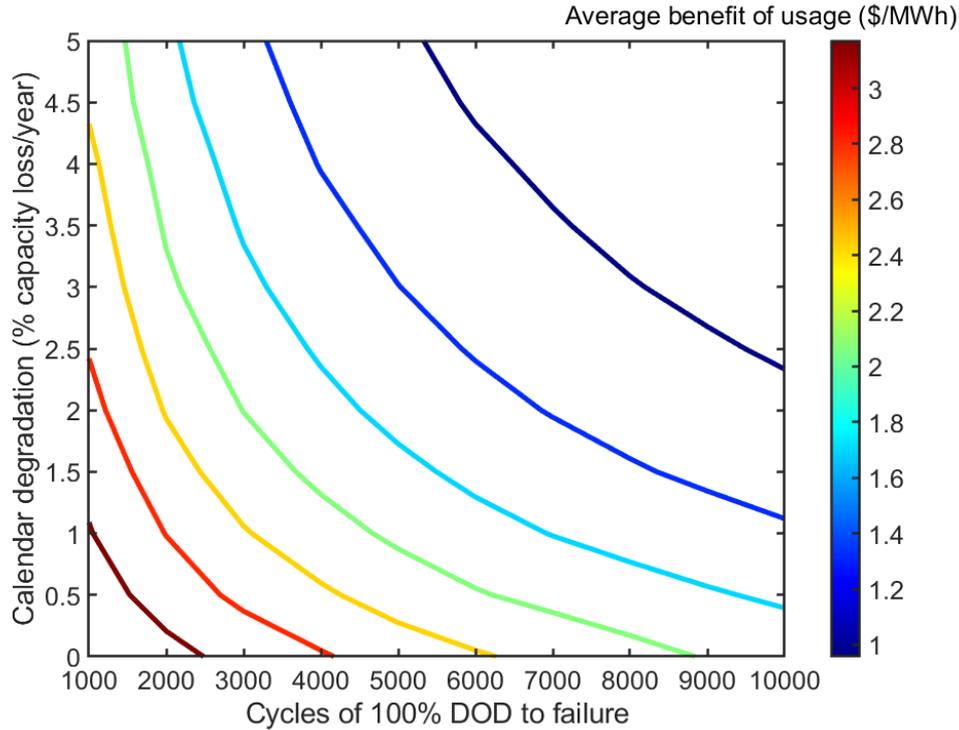

Figure 6. Sensitivity analysis on how the average benefit per unit of usage of EES arbitraging in California vary as the cycling and calendar degradation rates change

## 4 Discussions

In this paper, we define the economic EOL for EES, and illustrate its dominance over the physical EOL in some use cases. In general, if the profit opportunities over multiple years are essentially same, the annual profit of EES will decrease due to EES performance degradation – which means the system is less able to provide valuable services – and the temporal discounting of the EES owner. The net annual profit could become negative as the revenue cannot compensate for fixed O&M cost, and this is when the economic life of an EES asset ends. In the case study with a lithium-ion EES arbitraging in California energy markets, the annual profitability decreases by 20% after the capacity decrease to 70% and the impedance doubles. For utility-scale, commercial, and residential EES system, the economic lives are 8 years, 5 years, and 1 year, respectively. The economic life of EES varies with EES size, application and degradation characteristics, so the EES degradation characteristics should be carefully investigated and compared among different EES types when planning a project with specific business model or life requirement.

The existence of the economic life of EES could change how the energy storage research community views the useful life of EES and what to do at end of life, and in turn, the way to plan and deploy the EES. If the EES owners ignore the economic criterion when planning their system, it is likely that the EES will have to be replaced ahead of the initial schedule.

After the physical or economic EOL, EES may still be able to provide some services that require less cycling capability, like contingency reserve, back-up and black-start sources. If there is a secondary-use value or a second-hand market for EES, the economic EOL will tend to come even earlier: EES with higher SOH should be assessed or sold at a higher price, and thus selling the EES earlier may be more profitable. By considering economic EOL, the secondary-use value accounts for higher proportion of the total value of EES in some cases, as the EES tends to enter the secondary-use service earlier.

The sensitivity analysis informs EES scientists and owners that improving cycling capability is a less efficient way to extend the economic life and enhance the economic viability of EES when the calendar degradation rate is high, and vice versa. This provides important implication for EES R&D and planning to make trade-offs between cycling capability and calendar degradation rate.

## Acknowledgement

This work was partially supported by the US Department of Energy under Grant DEEE0007165.